\documentclass[journal=accacs,manuscript=article]{achemso}
\SectionNumbersOn
\usepackage[version=3]{mhchem} 

\graphicspath{ {./Figures/} }
\author{Michelle A. Hunter}
\author{Julia M. T. A. Fischer}
\author{Marlies Hankel}
\author{Qinghong Yuan}
\author{Debra J. Searles}
\email{d.bernhardt@uq.edu.au}
\affiliation[CTCMS]
{Centre for Theoretical and Computational Molecular Science, The Australian Institute for Bioengineering and Nanotechology, The University of Queensland, Queensland 4072, Australia}
\alsoaffiliation[SCMB]
{School of Chemistry and Molecular Biosciences, The University of Queensland, Queensland 4072, Australia}

\title[An \textsf{achemso} demo]
  {Doping Effects on the Performance of Paired Metal Catalysts for the Hydrogen Evolution Reaction }

\keywords{electrocatalysis, defects, hydrogen evolution reaction, doping effects, transition metal catalysts, graphene}

\begin{document}
\begin{abstract}
Metal heteroatoms dispersed in nitrogen-doped graphene display promising catalytic activity for fuel cell reactions such as the hydrogen evolution reaction (HER). Here we explore the  effects of dopant concentration on the synergistic catalytic behaviour of a paired metal atom active site comprised of Co and Pt atoms.  The metals are coordinated to six atoms in a vacancy of N-doped graphene. We find that HER activity is enhanced with increasing N concentration, where the free energy of  hydrogen atom adsorption ranges from 0.23 to -0.42 eV as the doping changes from a single N atom doped in the pore, to fully doped coordination sites. The results indicated that the effect of N is to make the Co atom more active towards H adsorption and presents a means through which transition metals can be modified to make more effective and sustainable fuel cell catalysts. 
\end{abstract}

\section{Introduction}
Rising global energy demands along with heightened environmental pressures necessitate a sustainable and cost effective method for the production of energy. The hydrogen fuel cell, which generates energy via the conversion of oxygen and hydrogen into water, is a promising, environmentally-friendly solution to this contemporary issue. Though a thermodynamically spontaneous reaction, large-scale use of the hydrogen fuel cell is limited by the Pt-based electrode materials required for the efficient catalysis of its half-cell reactions. Equally as important to consider are the reactions of the water electrolysis cell, notably the hydrogen evolution reaction (HER), and the oxygen evolution reaction (OER), as the sustainability of the hydrogen economy requires the hydrogen itself to be generated from water as opposed to hydrocarbon sources. Similarly to the fuel cell, the electrode materials required for the water electrolysis cell are metal-rich and are based on RuO$_{2}$ or IrO$_{2}$ for the OER at the fuel anode, and Pt/C for the HER at the cathode. \cite{Trasatti1980,Lee2012,norskovjk2005} It is therefore apparent that the amount of non-precious metal should be reduced in order for the hydrogen fuel cell to be viable. 

Metal atoms dispersed on graphene, either embedded or anchored to the surface, have been demonstrated to exhibit high activity towards the electrochemical reactions of fuel cells such as the HER. \cite{liu2017,zhangy2018,Bayatsarmadi2017,chengn2016} Commonly referred to as single atom catalysts, they are advantageous due to their low metal content, making them cheaper and more sustainable for use on an industrial scale.\cite{yangxf2013} Furthermore, the low metal coordination numbers and high surface area of the materials affords transition metals, which traditionally perform worse than precious metals, with the ability to perform comparably to such systems. As a result, the improvement of atomically-dispersed transition metals is an active area of study. \cite{qiuhj2015,Morozan2015,zhangl2016} Though some transition metal catalysts perform at the Pt/C benchmark, only a few systematic studies have been undertaken to determine the specific features of catalysts which underpin activity. \cite{Sun2018,Wang2017,zhangy2018}  For example, dopants are a means to easily alter metal coordination environment and the electronic properties of the materials, where N, B and S are often used. It would hence be desirable to develop a rationale for the design of catalysts. 

We recently studied the catalytic activity of pairs of atomic Co and Pt embdedded in N-doped graphene for the oxygen reduction reaction, the cathode reaction of the hydrogen fuel cell. \cite{Zhang2018} In unpublished work, the same materials outperformed Pt/C for the HER. Experimentally, single metal atoms of Co and Pt were found to be embedded 2-4 {\AA} apart which exhibited catalytic activity higher than state-of-the-art Pt/C under both acidic and basic conditions.\cite{Zhang2018, alec2018} Supported by our theoretical work, it was suggested that a synergistic effect between the metal atoms was responsible for the enhanced catalytic activity. This conclusion was drawn from models based on metal pairs of Co and Pt embedded N-doped vacancies in graphene which were developed from experimentally-observed distributions of inter-metal distances. \cite{Zhang2018} It was found that N doping was required to prevent the clustering of Pt atoms, and hence to increase their adhesion to the substrate.  However, such an effect was not observed for the Co species.\cite{Zhang2018} Another study of Co N-doped graphene demonstrated that pyridinic nitrogen was required for the enhancement of Co-based catalysts for the HER and was the most effective type of N for the coordination of Co.\cite{zhangy2018} Furthermore, it was found that experimentally, HER activity was not strictly dependent on the amount of N doping, but rather the geometries and coordination structures formed. \cite{zhangy2018} Another study of Co-N doping examined the effect of doping computationally. \cite{Wang2017} By altering the doping of CoN$_4$ sites, they found that hydrogen adsorption increased with N concentration. 

In this manuscript we present the results of a study of one of the pore models examined in the work by Zhang et al. \cite{Fischer2018,Zhang2018} for the HER. One of the most active systems in that work was found to be a paired cluster system, referred to as CoPt@N6V4, where the general notation M$_1$M$_2$@NXVY is used with M$_1$ and M$_2$ the metal atoms, X the number of nitrogen atoms, N, and Y is the number of carbon atoms removed to form a vacancy, V. The metal distance in this pore was the most common inter-metal distance observed experimentally ($\approx$2.5{\AA}) and hence it is interesting to study this material to understand this system.  The proximity of the metal atoms means it is more appropriate to be considered to contain a metal dimer as opposed to single atoms. The close proximity of the metal atoms also means that the active site involves bridging between the metal atoms and is highly dependent on the coordination environment provided by the surrounding C or N atoms, which depends on the N concentration. Therefore this is an ideal system to study the effect of N doping on Co, as well as Pt. Since the metal atoms and pore system will be the same in this case, the doped systems will be referred to as NX, with X denoting the number of nitrogen atoms. In this paper, we study the effect of pore doping on the performance of the paired metal cluster to observe how concentration and symmetry of doping affects the activity of the metal active site by increasing the doping concentration and determining the most stable dopant symmetry (X = 1-6). We examine the relationship between the stable doping configurations and observed HER activity. 

\section{Methods}
\subsection{Calculating HER Efficiency}\label{herintromech}
The HER is defined as: 
\begin{equation}\label{herstep1}
\mathrm{2H^{+}} + \mathrm{2{e^-}} \longrightarrow \mathrm{H_{2}},
\end{equation}
where protons from solution form gaseous hydrogen in the presence of a catalyst. The mechanism involves adsorption followed by two possible pathways to form H$_2$:
\begin{equation} 
\textrm{Step 1}: \hspace{0.5cm} \mathrm{H^+} + \mathrm{e^-} \longrightarrow \mathrm{{H^*}} 
\end{equation}
\begin{equation}\label{herstep2a}
\textrm{Step 2a}: \hspace{0.5cm} \mathrm {2H^{*}} \longrightarrow \mathrm{H_2} 
\end{equation}
\begin{equation}\label{herstep2b}
\textrm{Step 2b}: \hspace{0.5cm} \mathrm{H^*} + \mathrm {e^-} + \mathrm {H^+} \longrightarrow \mathrm {H_2}. 
\end{equation}
In both mechanisms, the critical step has been established to be the adsorption of the hydrogen atom onto the electrode surface, denoted H$^*$ which is featured in Step 1.\cite{norskovjk2005} Irrespective of whether the second H atom is bound to an adjacent site (Step 2a) or from the solution (Step 2b), the thermodynamically and kinetically hindering step is adsorption and desorption. As the HER is referenced to the reversible hydrogen electrode (RHE), the reaction profile is the same in both directions.  This means that the ideal behaviour for the HER is when $\Delta{G_\mathrm{H^*}}$ is defined as reversible (i.e. $\Delta{G_\mathrm{H^*}}$ = 0). 

Catalytic activity of the materials was gauged using $\Delta{G_\mathrm{H^*}}$, defined as the free energy of hydrogen adsorption. This has found to be correlated with the exchange current density, which is proportional to the number of H$_2$ produced per unit surface area.\cite{norskovjk2005}  This criterion places Pt/C, modelled as Pt(111), which has a $\Delta{G_\mathrm{H^*}}$ = -0.09 eV, at the top of volcano plots of $\Delta{G_\mathrm{H^*}}$ against current density as a measure of activity. This is because the equilibrium potential, $U_\mathrm{0}$, is defined as 0 V in the computational hydrogen electrode model (see Supplementary Information). \cite{norskovjk2005} As the kinetic barriers of proton transfer to the surface of electrodes have been found to be comparable to those in water, it is appropriate to consider the binding of H$^*$ on a thermodynamic basis.\cite{Tripkovic2010} For a given coverage of H (number of H atoms) per unit cell, the reaction energy of binding is given by: 
\begin{equation}
\Delta{E_\mathrm{nH^*}} = \frac{E_\mathrm{SF+nH} - E_\mathrm{SF} - \frac{\mathrm{n}}{2}E{_\mathrm{H{_2}}}}{\mathrm{n}} 
\end{equation}
where n is the number of H atoms adsorbed on the surface. We use the combined ZPE and entropy correction of 0.24 eV from N\o rskov et al.\cite{norskovjk2005} giving the free energy of adsorption as :  
\begin{equation}
\Delta{G_\mathrm{H^*}} = \Delta{E_\mathrm{H^*}} + 0.24 \hspace{0.2cm} \mathrm{eV}.  
\end{equation}

To improve on the performance observed for Pt(111), it is apparent that any new system should have $\Delta{G_\mathrm{H^*}}$ $\leq$ $\mid$ 0.09 $\mid$ eV. The energy can be positive or negative as the sign will dictate which step of the two-part reaction is thermodynamically limiting - the adsorption of hydrogen atoms or the desorption of the hydrogen molecule. 

\subsection{Computational Parameters}
All calculations were performed within the density functional theory framework using the Vienna Ab initio Simulation Package (VASP). \cite{kresseg1996} The revised Perdew-Burke-Ernzerhof functional (RPBE) with the DFT-D3 dispersion correction and spin polarisation were employed.\cite{grimmes2010, hammerb1999} A periodically-projected plane wave basis set was employed using the projector-augmented wave (PAW) method with a kinetic energy cutoff of 450 eV. \cite{kresseg1999}  The Brillouin zone was sampled using a 2 x 2 x 1 $\Gamma$-centred $k$-point mesh. An 8 x 4 graphene supercell was used, with lattice dimensions 19.74 x 17.09 {\AA} in the basal plane and 20 {\AA} perpendicular to ensure a sufficiently large vacuum. The optimal lattice parameters, supercell size, and defective N-doped graphene models were designed and tested using a single graphene sheet. 

For the visualisation of the coordinate data, the program VESTA was employed.\cite{Momma:db5098} Bader charge analysis was conducted using the Bader program developed by the Henkelman group. \cite{ISI:000262897400005}

The formation energies ($E_\mathrm{f}$) of the surfaces were calculated using the following:

\begin{equation}
E_\mathrm{f} = E_\mathrm{CoPt@NXVY} - E_\mathrm{GR} + (X+Y){\mu_\mathrm{C}} - X\mu_\mathrm{N} - E_\mathrm{Co} - E_\mathrm{Pt},
\end{equation}

where $E_\mathrm{CoPt@NXVY}$ is the total energy of the surface and metal atoms, $E_\mathrm{GR}$ is the energy of perfect graphene, ${\mu_\mathrm{C}}$ is the chemical potential of a C atom in perfect graphene, $\mu_\mathrm{N}$ is the chemical potential of N in N$_{2}$, $E_{Co}$ is the energy of the Co atom, $E_{Pt}$ is the energy of the Pt atom and $X$ and $Y$ are the number of carbon atoms removed to form a vacancy and the number of nitrogen atoms for doping, respectively. In this manuscript $Y=4$ in all cases. The relative binding energies of the metal atoms to the pores were calculated by referencing to the undoped carbon material embedded metal cluster, CoPt@N0V4: ($\Delta E_\mathrm{b}=E_f(CoPt@NXV4)-E_f(CoPt@N0V4)$)  (see Supplementary Information). 

\section{Results \& Discussion}
\subsection{Surface Properties}
Paired metal systems were studied with different doping concentrations (X = 1-6) and by studying possible configurations for each concentration to determine the most thermodynamically stable configuration for each concentration, X. 

\begin{figure}[!htb]
\centering
\includegraphics[scale=0.95]{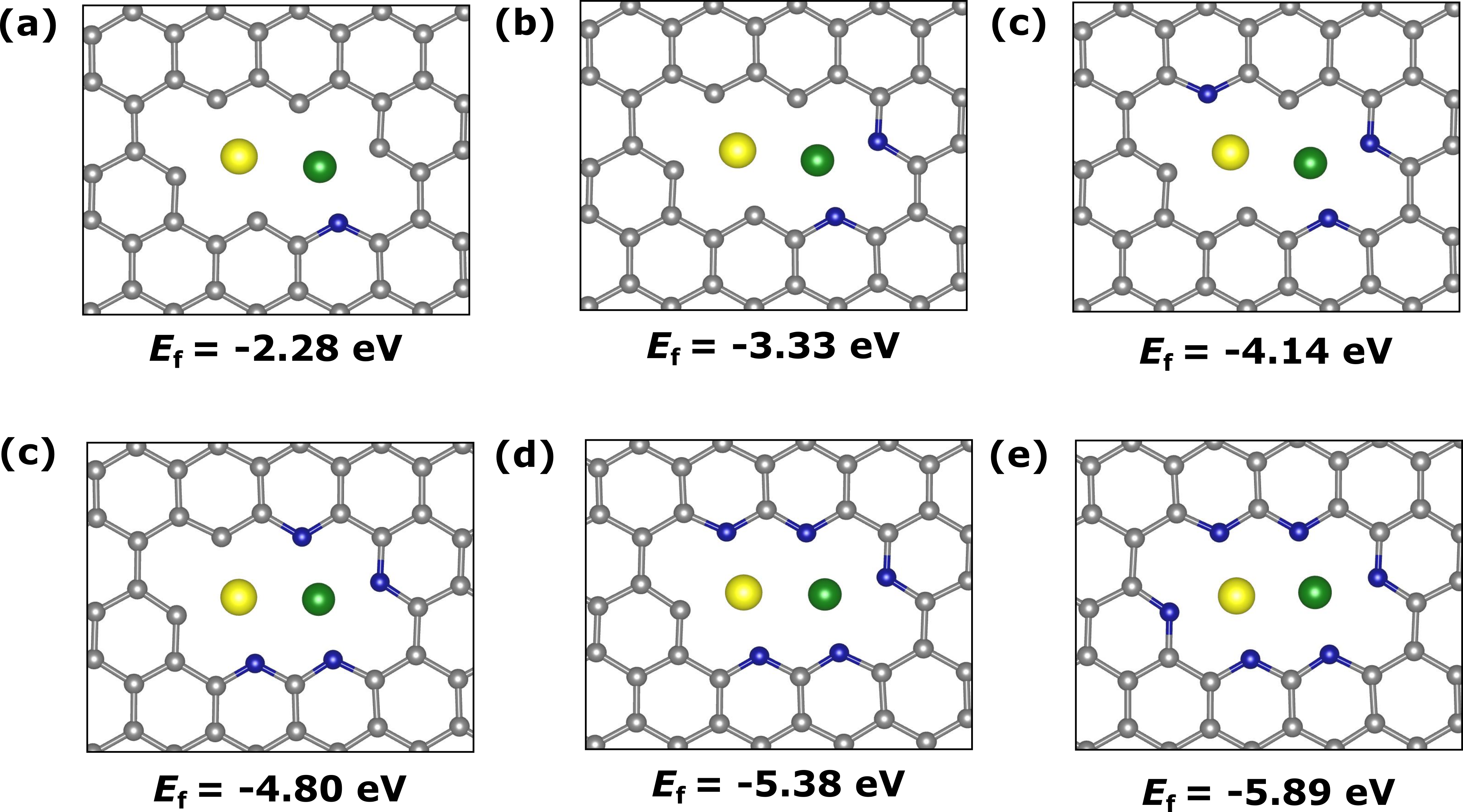}
\caption{The doped graphitic surfaces considered. The most stable configurations for the materials of the form CoPt@NXV4 (X = 1-6) are presented. The system for X = 0 was also tested and is not shown since there are no isomeric forms with this geometry (see Supplementary Information). The formation energies $E_\mathrm{f}$ for the systems are displayed in eV. Atom colours: C in grey, N in blue, Co in green and Pt in yellow. The Co-Pt bond-lengths range from 2.36 - 2.43 {\AA}.}
\label{SF}
\end{figure}

\begin{figure}[!htb]
\centering
\includegraphics[scale=1]{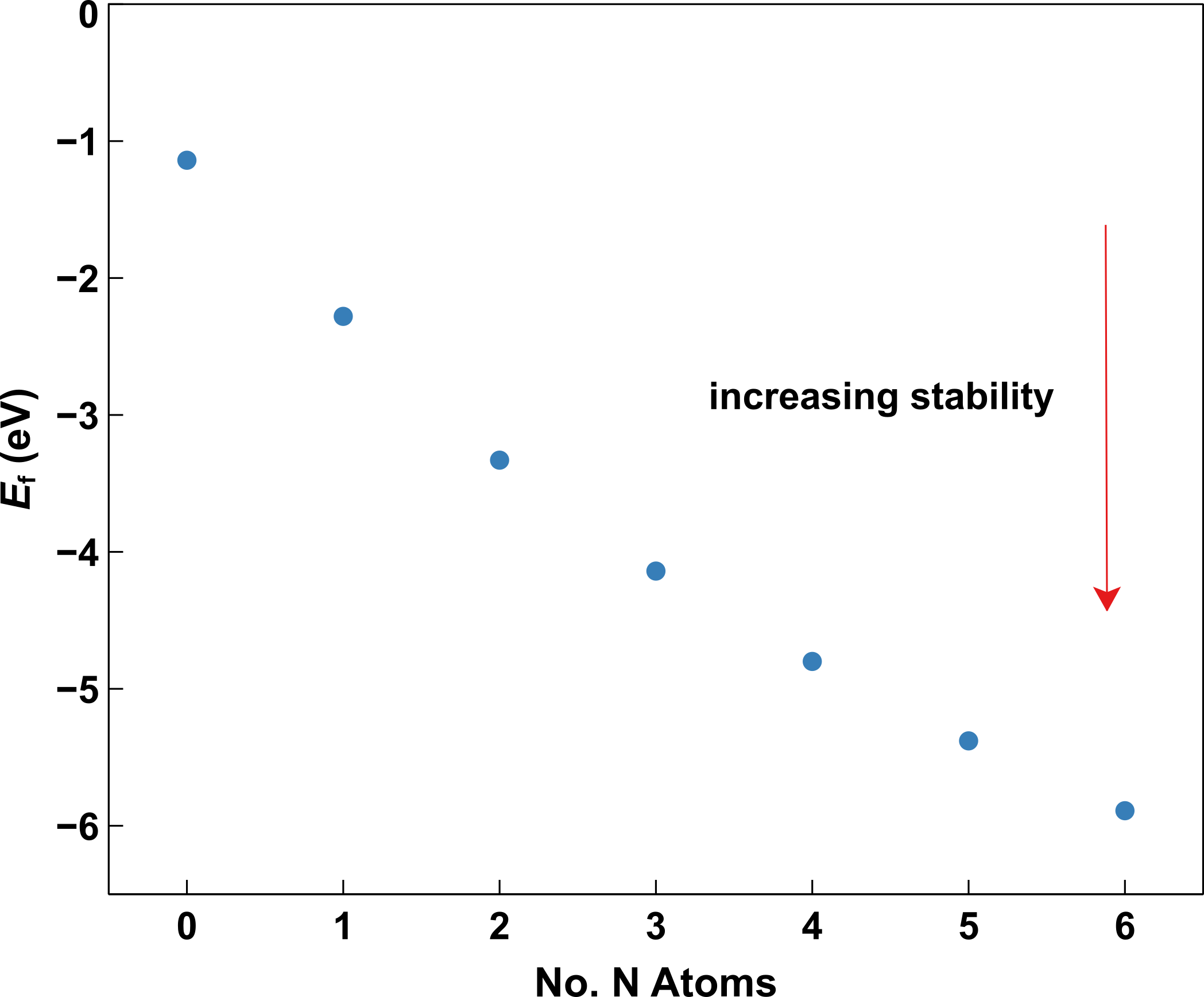}
\caption{Formation energy ($E_\mathrm{f}$) of the CoPt@NXV4 systems as a function of increasing N dopant concentration. As the values become more negative, the systems become more thermodynamically stable.}
\label{Ef}
\end{figure}

In Figure \ref{SF}, the most stable configurations of the pores and their formation energies ($E_\mathrm{f}$) are shown. The trend in $E_\mathrm{f}$ shown in Figure \ref{Ef} demonstrates that the surfaces become more thermodynamically stable with the amount of N (see Figure \ref{Ef}). This indicates that the amount of dopant in the most stable system is the $X=6$. Furthermore, the metals are embedded in the graphene pores at distances from 2.36-2.43 {\AA} apart (see Table S1). The stability of these systems indicate that they effectively form materials where the metal cluster is embedded in a doped graphene pore and which are likely to bind intermediates in a concerted reaction. It is worthwhile noting that the metal-metal bond-length does not increase linearly with doping concentration, rather is 2.42 {\AA} for N1 and decreases with doping to N3 (2.40 {\AA}), increases again for N4 and decreases to 2.36 {\AA}, which is the shortest bond-length, for N6. In Table S1, it appears that there is a correlation between $\mu_\mathrm{Co}$, the magnetic moment on the Co atom, and the observed cluster bond-length, which are directly proportional. It is interesting to note that although there is only an increase in the number of N bound to Pt from N = 4 and above (and not the number bound to Co), the properties of the Co atom still change considerably. This suggests that the activity of Co changes accordingly. Therefore, it appears that the changes in the geometry of the pore due to coordination to N rather then C are associated with a change in the electronic properties of the metal atoms. From the different structures observed, it appears that the most stable structure of the form NXV4 in each case is the structure of N(X-1)V4 with substitution of an extra C atom with a N atom, with the exception of the structure N4V4.  

The difference of the binding energies of the metal clusters on the N-doped pores and that of the undoped pore were calculated to elucidate the effect of dopant concentration on the cluster binding strength (Table S1). In this case, a more positive value indicates that the metals bind to the doped pore less strongly than they do to the undoped pore. The trend in relative binding energy increases from 0.05 eV for N1, to 5.55 eV in N6. The relative binding energy increases almost linearly with dopant concentration until N3, at which point there is a large increase in the binding energy for the system in going from N3 to N4 (Table S1). It appears that the N weakens the metal cluster interactions with the pore ($\Delta E_b$), whilst increasing the metal-metal interactions.  This is consistent with the observed metal cluster bond-lengths. These observations based on the formation energy and the binding energy of the systems demonstrate that the substitution of C with pyridinic N is thermodynamically favourable in each case, whereas changing coordination from C to N results in weakening of the cluster interaction with the doped pore. Based on our experimental results, it is likely that this weakening of the interaction is due to the weakening of the Co interaction of the pore, since N was required to promote Pt adhesion.\cite{Zhang2018}

\subsection{Hydrogen Adsorption}
\begin{figure}[!htb]
\centering
\includegraphics[scale=0.65]{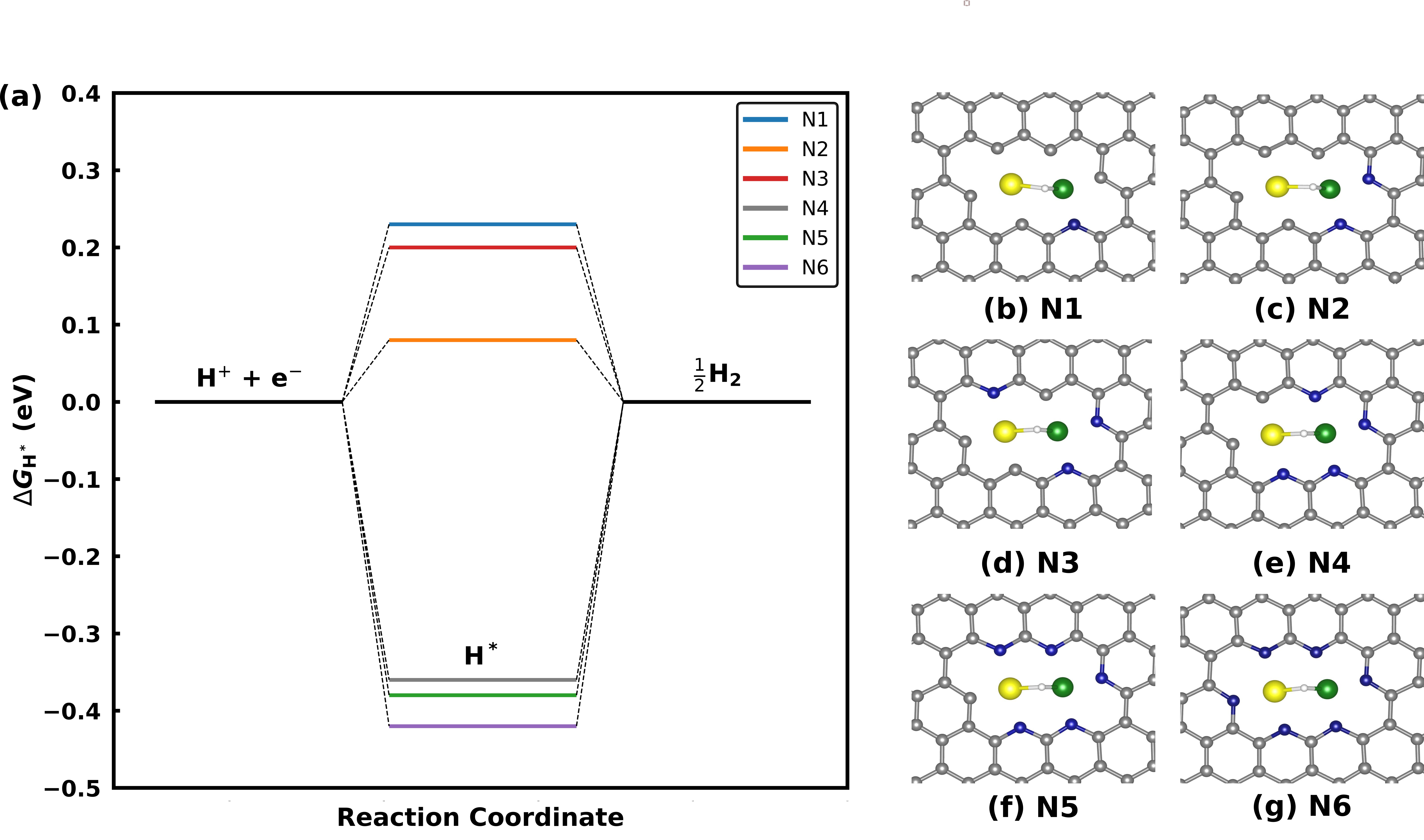} 
\caption{Free energy profile for the adsorption of a hydrogen atom on the doped catalytic surfaces. NX, where X is the number, denotes the number of nitrogen atoms featured in the doped pore. The free energy profile is shown at the equilibrium potential for the HER. In cases where the hydrogen atom prefers to bind to the carbon atoms (N1 and N3), the results are for the most stable site involving binding to the metal atoms.}
\label{freeG}
\end{figure}

The efficiency of the HER was evaluated on the most stable surfaces by calculating $\Delta{G_\mathrm{H^*}}$ for a single H. The free energy profile for hydrogen adsorption is shown in Figure \ref{freeG}, where the HER is defined to be reversible. The optimised geometries for H adsorption to the bridge site between the metals are also shown. It can be seen that the preferred site for H adsorption is between the metal atoms embedded in the pore, except for the systems N1 and N3 where the most stable adsorption sites were to coordinating carbon atoms which also occurs when the pore is undoped (see Figures S1 and S2). 

For the doped systems, in most cases the H adsorption strength increases with N coordination. The exception is for the N3 pore, which has a higher adsorption free energy than N2. From the free energies, it appears that the substitution of a carbon atom and its coordination to Pt makes the bridge position for the metal cluster more weakly interacting with the H atom than with N3 which is non-spontaneous. For N1-N3, the adsorption step is non-spontaneous as the values are positive, and the reverse is true for N4-N6. These findings suggest that the weakening of metal-pore bonds is required for an effective interaction with the H atom. Furthermore, the less stable interaction that occurs at the bridge site for N1 and N3 is consistent with N being required to make Co more active for H adsorption which occurs for the other pores, as N4 and above have Co coordinated to three N atoms. The C atoms around the pore are otherwise under-coordinated and therefore would be expected to be comparatively active towards H adsorption. The preferential adsorption of H to C in N3 and to the cluster in N2 suggests that the coordination of Pt to N weakens the interaction between the cluster and the H atom, or conversely makes the surrounding C atoms more reactive. This would also explain why the N3 exhibits a weaker adsorption energy than N2. 

To test whether the coordination of Pt to N destabilises interactions at the bridge site for N3, H was also adsorbed to CoPt embedded in the N3 pore where Co was coordinated only to N (see Figure S3). Adsorption at the bridge site had a $\Delta{G_{H^*}}$ = -0.29 eV and was overall the most stable adsorption site. Therefore, the coordination of the first N atom to Pt in exchange for C changes the adsorption energy of the cluster site by -0.49 eV. This difference in energy demonstrates that the adsorption energy is greatly influenced by the coordination of Co. For the systems N4-N6, the nitrogen coordination of Pt increases by 1. The effect of this doping is a small increase in the hydrogen adsorption strength to the metal dimer where $\Delta{G_\mathrm{H^*}}$ decreases from -0.36 eV to -0.42 eV.  Despite the local magnetic moments on Pt not changing significantly , it appears that the coordination of Pt affects the Co atom significantly (see Table S3). This is likely due to the adjustment of the Pt to accommodate for Pt-N bond-lengths which is observed in the metal-H bond-lengths in Table S2.                            

Furthermore, it is interesting to note that hydrogen adsorption activity can be correlated with the cluster bond-lengths observed in the free surface and subsequently the magnetic moments on the metal atoms before H adsorption. It appears that since the cluster bond-length in the N2 surface is shorter than the N3 bond-length, that this may be the reason for the stronger H binding observed. This indicates that it is the coordination changes which accompany the N doping that are responsible for the observed activity. This also explains the change in doping from N3 to N4 which shows that N doping of Co to three N bonds increases adsorption strength by 0.56 eV. Hence, the binding strength of the H atom appears to be much more significantly affected by the N coordination of Co, whereas increasing the coordination to Pt for structures results in minor changes from 0.2-0.3 meV. Therefore, the gap in the behaviour between N1-N3 and N4-N6 is a result of the coordination of Co to the third N atom which is consistent for the stable surfaces from N4-N6 and the addition N3 structure tested for H adsorption. Furthermore, the Co-Pt bond-length also increases in N1-N6 from 2.43-2.47 {\AA}. This is likely because the coordination of H to the Pt atom is increasing, and the bond between the metal atoms weakens.  

\subsection{Charge Analysis \& Magnetism}
To analyse the effects on the local charge arising from the N doping, a Bader charge analysis was performed on the system with H adsorbed. 

\begin{table}[!htb]
\caption{Bader charges, $Q$, on the paired metal atoms for the materials with different doping concentrations.}
\centering
\begin{tabular}{|c|c|c|c|}
\hline
System	&$Q_{Co} / e$	&$Q_{Pt} / e$	&$Q_{M1 + M1} / e$\\ \hline
N1V4	&0.72	&0.31 &1.02\\ \hline
N2V4	&0.76	&0.30 &1.06\\ \hline
N3V4	&0.79	&0.37 &1.16\\ \hline
N4V4	&0.83	&0.34 &1.17\\ \hline
N5V4	&0.82	&0.44 &1.26\\ \hline
N6V4	&0.78	&0.51 &1.29\\ \hline
\end{tabular}
\label{baderH}
\end{table}

From the Bader analysis, the total charge on the metal atom pair decreases with the degree of N doping from 1.09 to 1.29$e$. Therefore, the increase in doping to N6 enables 0.20$e$ to be lost from the metal cluster. This indicated that there is a loss of electrons from the metal atom cluster consistent with an increased binding strength of the H atom to the metal cluster at the higher concentrations. It therefore appears that the metal atoms are able to more readily oxidise with the increased dopant, resulting in the stronger binding observed. However, examining the individual metal atoms, it appears that charge loss is occurring on both the Co and Pt atoms, where the result is generally an increase in the Co-H bond-length as the Pt atom coordinates more strongly to the H atom (Table S1). These changes are consistent with Pt forming a stronger bond with the adsorbed H atom as the Pt-H bond-lengths get progressively shorter. These findings are consistent with the idea that the weakened bonds with the pore structure that occurs with doping enables the metal cluster, in particular, Co, to lose electrons more readily as demonstrated by the Bader charge as well as the observed changes in magnetic moments (Table S3).  

\section{Conclusions \& Outlook}
Based on the study undertaken, nitrogen doping of graphene vacancies is a means to make transition metals, such as Co, more active towards H adsorption and hence improve catalytic activity towards the HER.  It appears that the effect of N doping is to predominantly weaken the metal-substrate interactions which in turn make the active site more effective. The calculated formation energies indicate that the doping of graphene vacancies with pyridinic nitrogen is highly thermodynamically favourable, supporting the proposal that Co-Pt clustered systems should be highly active towards hydrogen adsorption. This was demonstrated by the behaviour observed for the N4-N6 systems. It is found that shorter cluster bond-lengths result in stronger binding, whilst a nitrogen coordination number of three is required for the adsorption step of the HER to be spontaneous. Furthermore, the work demonstrated that N2 system enables the catalytic system studied to perform comparably to the industrial benchmark, Pt/C, with a $\Delta{G_\mathrm{H^*}}$ of 0.08 eV. The findings of this study demonstrate the versatility of combining difference dopant and metal atoms for specific catalytic reactions. It exemplifies the wide applicability of dispersed metal atoms for the catalysis of fuel cell reactions.  

\begin{acknowledgement}

The authors thank the Australian Research Council for support of this project through the LIEF program. We acknowledge access to computational resources at the NCI National Facility through the National Computational Merit Allocation Scheme supported by the Australian Government. This work was also supported by resources provided by the Pawsey Supercomputing Centre with funding from the Australian Government and the Government of Western Australia. We also acknowledge support from the Queensland Cyber Infrastructure Foundation (QCIF) and the University of Queensland Research Computing Centre. 

\end{acknowledgement}

\begin{suppinfo}

Figures detailing additional information about the investigated surfaces and their properties, and the methods employed are available. This material is available free of charge. 

\end{suppinfo}

\bibliography{biblio}


\end{document}



\section{Supporting Information}
\renewcommand{\thefigure}{S\arabic{figure}}
\renewcommand{\thetable}{S\arabic{table}}
\renewcommand{\theequation}{S\arabic{equation}}

Throughout this work the computational hydrogen electrode model (CHE) was employed by equating each electron-proton pair as half that of gaseous hydrogen at equilibrium and hence using the reversible hydrogen electrode as the reference potential.\cite{norskovjk2004} This simulates the effect of a potential on the reaction mechanism by considering the reaction barriers of intermediates interacting with the surface as the primary source of potential dependence. Furthermore, the effect of the potential on the free energy of each intermediate is defined as follows: 
\begin{equation}
\Delta{G} = \Delta{E} + {\mathrm \Delta{ZPE}} - {T{{\Delta{S}}}} - \Delta{G_\mathrm{{pH}}} - neU,
\end{equation}
where $\Delta{E}$ is the DFT-calculated reaction energy, $\Delta{ZPE}$  the zero-point energy, $T\Delta{S}$, the entropic correction, ${\Delta{G_\mathrm{{pH}}}}$ is the effect of the pH on the potential and ${neU}$ describes the potential $eU$ relative to the reversible hydrogen electrode scaled by the number of electrons, n. \cite{norskovjk2004}

The relative binding energies ($\Delta E_\mathrm{b}$) of the the doped systems were calculated by referencing to the undoped carbon pore (CoPt@N0V4):

\begin{equation}
\Delta E_\mathrm{b} = E_\mathrm{CoPt@NXV4} - E_\mathrm{NXV4} - (E_\mathrm{CoPt@N0V4} - E_\mathrm{N0V4}),
\end{equation}
where $E_\mathrm{CoPt@NXV4}$ is the energy of the total doped system, $E_\mathrm{NXV4}$ is a single-point energy of the embedded pore with the metal atoms removed, $E_\mathrm{CoPt@N0V4}$ is the total energy of the undoped embedded metal system, and $E_\mathrm{N0V4}$ is the single-point energy of the undoped embedded system with the metals removed. 

\begin{table}[!htb]
\centering
\caption{Properties of the doped surfaces. Relative binding energies of the metal cluster ($\Delta E_\mathrm{b}$)  and formation energies ($E_\mathrm{f}$) of the doped surfaces, local magnetic moments on the metal atoms ($\mu_\mathrm{M}$) and metal cluster bond-lengths ($r_\mathrm{Co-Pt}$).}
\begin{tabular}{|c|c|c|c|c|c|}
\hline 
System	&$\Delta E_\mathrm{b} /$eV	&$E_\mathrm{f} /$eV	&$\mu_\mathrm{Co} / \mu_\mathrm{B}$	&$\mu_\mathrm{Pt} / \mu_\mathrm{B}$ & $r_\mathrm{Co-Pt} / \angstrom$\\ \hline
N1V4	&0.05	&-2.28	&1.16	&0.03  & 2.42 \\ \hline
N2V4	&0.78	&-3.33	&0.84	&0.01  & 2.39 \\ \hline
N3V4	&1.16	&-4.14	&0.41	&0.00  & 2.40 \\ \hline
N4V4	&3.75	&-4.80	&1.73	&0.19  & 2.43 \\ \hline
N5V4	&4.64	&-5.38	&1.62	&0.17  & 2.41  \\ \hline
N6V4	&5.55	&-5.89	&1.27	&0.06  & 2.36 \\ \hline
\end{tabular}
\end{table}

\begin{figure}[!htb]
\includegraphics[scale=0.3]{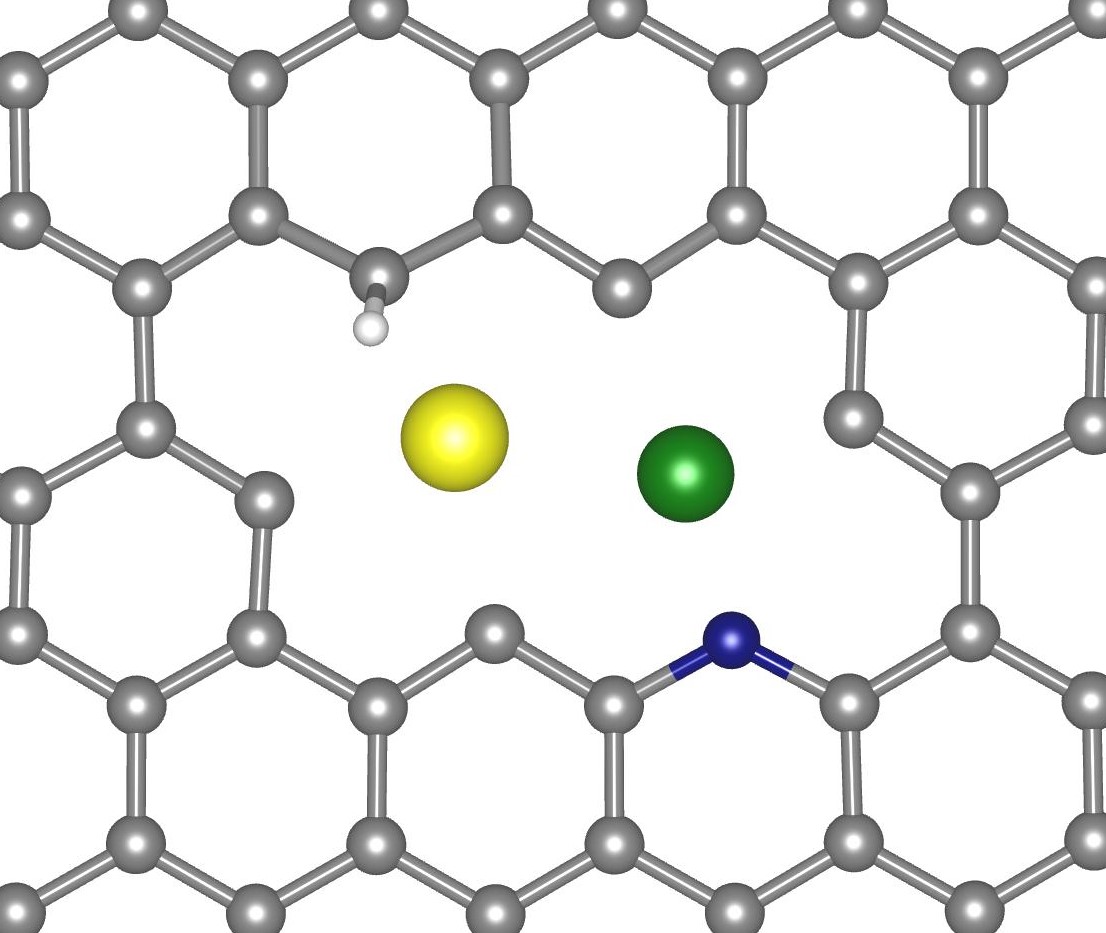}
\caption{Global minimum energy geometries for hydrogen adsorption to the system CoPt@N1V4. Colours: C, grey, N, blue, Pt, yellow, Co, green, and H in white. Calculated free energy of adsorption, $\Delta{G_\mathrm{H^*}}$, is -0.18 eV.} 
\label{CPN1V4}
\end{figure}

\begin{figure}[!htb]
\includegraphics[scale=0.3]{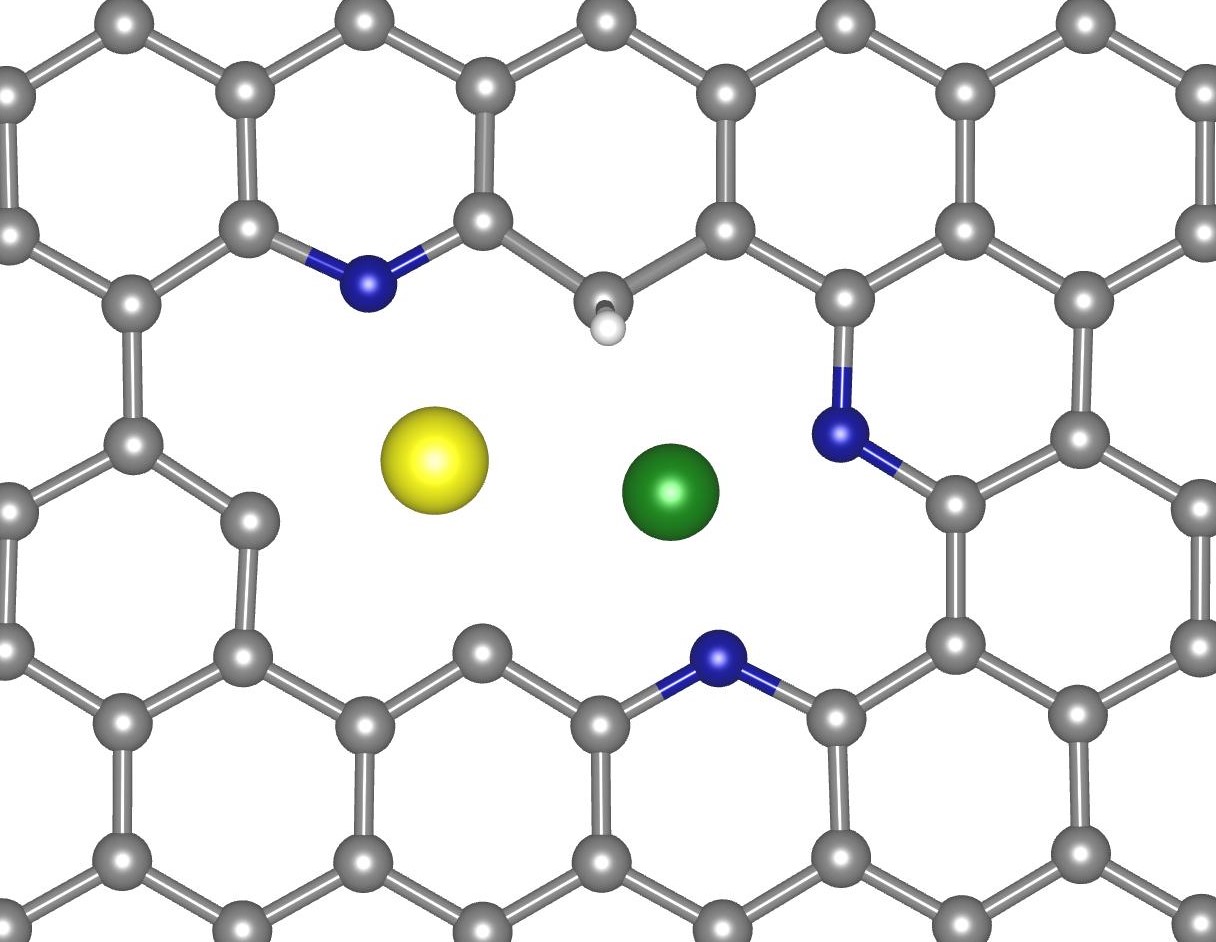}
\caption{Global minimum energy geometries for hydrogen adsorption to the system CoPt@N3V4. Colours: C, grey, N, blue, Pt, yellow, Co, green, and H in white.} 
\label{CPN1V4}
\end{figure}

\begin{figure}[!htb]
\includegraphics[scale=0.3]{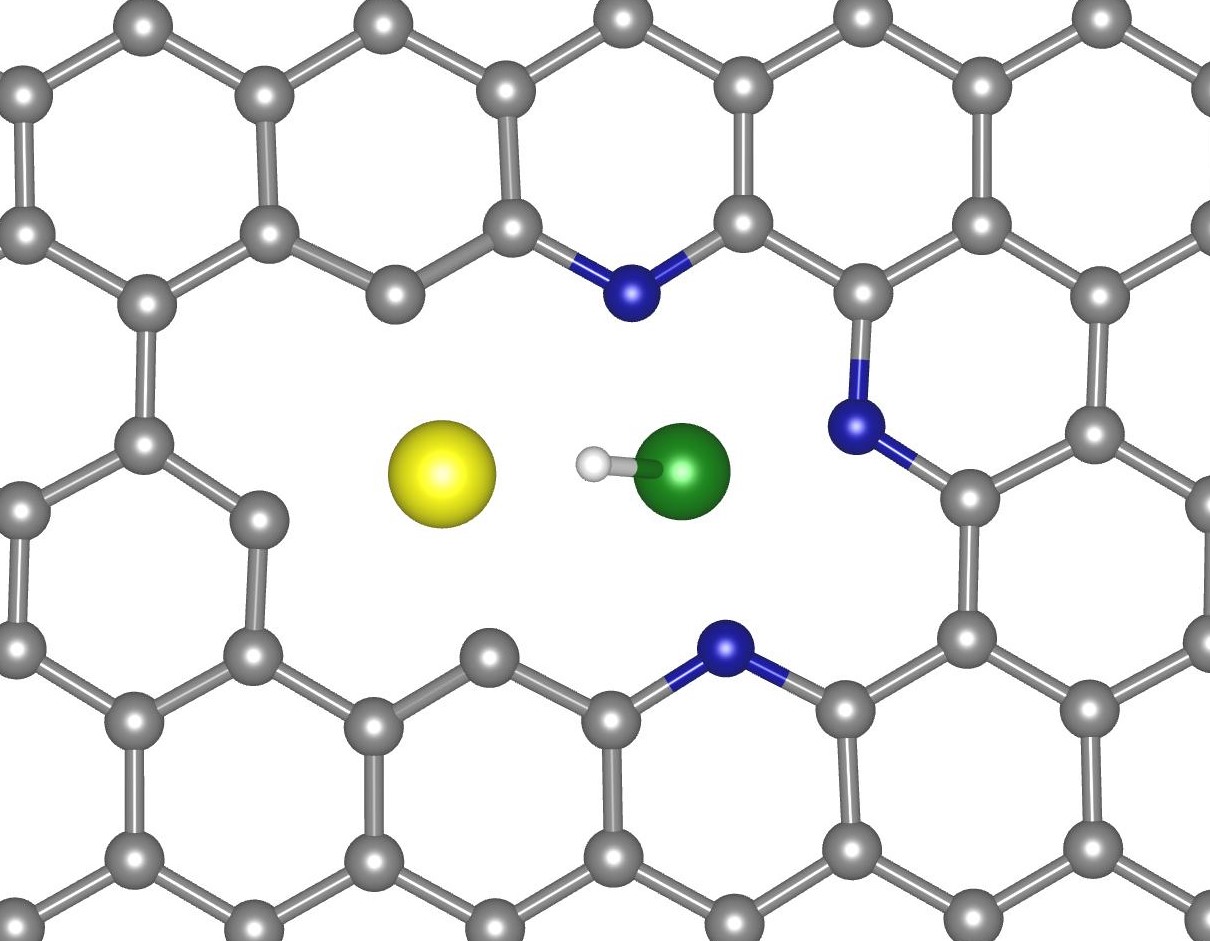}
\caption{Minimum energy bridging structure for the system, N3V4. Colours: C, grey, N, blue, Pt, yellow, Co, green, and H in white. } 
\label{CPN1V4}
\end{figure}

\begin{table}[!htb]
\centering
\caption{Metal-hydrogen bond-lengths and metal-metal bond-lengths for systems where a single hydrogen atom is adsorbed.}
\begin{tabular}{|c|c|c|c|}
\hline 
System	&$r_\mathrm{Co-H} / {\angstrom}$	&$r_\mathrm{Pt-H} /{\angstrom}$ &$r_\mathrm{Co-Pt} / \angstrom$\\ \hline
N1V4	&1.59	&1.89 & 2.43 \\ \hline
N2V4	&1.57	&1.84 & 2.46 \\ \hline
N3V4	&1.61	&1.77 & 2.46 \\ \hline
N4V4	&1.64	&1.75 & 2.49 \\ \hline
N5V4	&1.62	&1.74 & 2.47 \\ \hline
N6V4	&1.65	&1.66 & 2.47 \\ \hline
\end{tabular}
\end{table}

\begin{table}[!htb]
\centering
\caption{Free energies of H adsorption ($\Delta{G_\mathrm{H^*}}$) and the local magnetic moments on the metal atoms ($\mu_M$).}
\begin{tabular}{|c|c|c|c|}
\hline 
System	&$\Delta{G_\mathrm{H^*}} /$eV	&$\mu_\mathrm{Co} / \mu_\mathrm{B}$	&$\mu_\mathrm{Pt} / \mu_\mathrm{B}$\\ \hline
N1V4	&0.23	&0.55	&0.00\\ \hline
N2V4	&0.08	&0.00	&0.00\\ \hline
N3V4	&0.20	&1.13	&0.09\\ \hline
N4V4	&-0.36	&0.93	&0.10\\ \hline
N5V4	&-0.38	&0.82	&0.05\\ \hline
N6V4	&-0.42	&0.57	&0.03\\ \hline
\end{tabular}
\end{table}

\clearpage
\bibliography{biblio}
